# Proposed experiment to test the non-locality hypothesis in transient light-interference phenomena

**Masanori Sato**

*Honda Electronics Co., Ltd.,*

*20 Oyamazuka, Oiwa-cho, Toyohashi, Aichi 441-3193, Japan*

**Abstract** The transient phenomena of the Mach-Zender interferometer are discussed. To test the non-locality hypothesis, a single mode laser with a large coherence length is used. The behavior of a photon and its wave packets in the paths of the interferometer are discussed. Coherent photons have wave packets that overlap, thus their interference pattern is influenced by the overlap of the wave packets of other photons in transient phenomena. The proposed transient light-interference experiment will provide experimental data testing the non-locality hypothesis.

1. Introduction

Non-locality appears in the de Broglie-Bohm picture in the theory of hidden variables by Bohm [1, 2] as quantum potential. Bell [3] described the de Broglie-Bohm picture as attractive, which refers a photon as a particle guided by a wave. In this report, the wave packet interpretation is explained using the de Broglie-Bohm picture. The illustration shows a particle guided by a wave packet.

Delayed-choice experiments were first proposed by Wheeler [4] and experimentally examined by Hellmuth et al. [5]. In an experimental setup using a Mach-Zender interferometer, there arises the critical problem of causality in transient phenomena as described in section 2. Hellmuth et al. [5] carefully eliminated these critical conditions; however, I think it worthwhile to check these experimental conditions. This is one of the starting points of this proposal.

Interference seems to have similarities to quantum entanglement: Photons travel at the speed of light, $c$; however, the speed of interference does not seem to be restricted by the speed of light, $c$. There have been few experiments on transient phenomena in quantum mechanics. Transient phenomena like interference pattern formation, are of interest. Coherent phenomena show that the speed of interference pattern formation does not appear to be restricted by the speed of light, $c$.

In the calculation of two-slit interference experiments, an Airy pattern was clearly simulated [1, 2]. However, at this stage, I have not found any calculation of Young' double slit pattern, which is observed at larger distance. An Airy pattern is observed at the distance around 0.1~1 m; Young's double slit pattern is clearly observed at the distance of more than 1 m. That is, each Airy pattern splits into Young's double slit pattern, according to the distance from the double slit. At the near range (around 5 mm from the double slit), two bright lines are clearly observed, at the middle range (around 0.1~1 m from the double slit), an Airy pattern is observed: according to the distance, two bright lines are transformed into an Airy pattern. At far range (more than 1 m from the double slit), Young's double slit pattern is clearly observed: according to the distance, each Airy pattern is transformed to Young's double slit pattern. Therefore, at this stage, I think that the calculations were carried out for an Airy pattern; Young's double slit pattern has not been calculated yet.

In the two-slit experiment, when one of the two slits is closed, Young's double slit pattern is transformed to Airy pattern: therefore, transient phenomena from Young's double slit pattern to the Airy pattern will be observed. This experiment is very primitive; however, it arises from a very fundamental problem. In the transient phenomena, photons move transverse to the pattern of Young's double slit to transform to the Airy pattern. A photon travels only 3 mm, thus it takes $10^{-11}$ second. If the transformation of Yong's double slit to the Airy pattern occurs at 3 m from the double slit, it takes $10^{-8}$ seconds for a photon to travel from slit to screen. Thus, a photon transversely travels 3 mm at the speed of light, $c$, and the pattern transformation looks like a non-local phenomenon: the person who watches the pattern transformation knows the information of the two-slit (one slit opens or two slits open) non-locally. This paper started from the above discussion to obtain the transient phenomenon that proves non-locality.

Transient phenomena of Young's double slit experiment are rather difficult to observe experimentally. Thus, a Mach-Zender interference experiment is proposed.

In this study, discussions are carried out within the theoretical framework of the de Broglie-Bohm picture. I consider that the wave packet hypothesis is equivalent to Bohm's quantum potential. The photon is guided by a wave packet, i.e., a quantum potential. A photon is a local particle guided by a non-local wave. Local means that the "photon travels at the speed of light, $c$."

I show a new experimental procedure for observing transient phenomena and obtaining experimental evidence of non-locality.

2. Wave packet interpretation

In this section, I describe a wave packet interpretation of the interference pattern generation. **Figure 1** shows a Mach-Zender interferometer with acousto optic modulators (AOMs). After a photon and its associated wave packets have passed through the AOMs and before their arrival at beam splitter 2, if the AOMs are turned off, as shown in **Fig. 2**, the interference of the photon seems to be determined by the wave packets, which travel at the speed of light along the path regardless of the condition of the AOMs. Thus, the interference of the photon that has passed through an AOM is not controllable by the AOMs. In the experiment by Hellmuth et al. [5], these experimental conditions were purposely eliminated. Therefore, there are no experimental data on the behavior of a photon that has passed through the AOMs before its arrival at beam splitter 2 by controlling the AOMs.

This conclusion differs from the interpretation in which the setting of experimental setup at the time when a photon reaches beam splitter 2 defines the interference condition. That is, in **Fig. 2**, the experimental setup does not appear to cause interference. However, according to wave packet interpretation, interference occurs at beam splitter 2.

At this stage, discussions are carried out using wave packet interpretation.

## 3. Experimental setup for analyzing transient interference phenomena

Now, let us discuss a coherent experiment using a single mode laser with a coherent length of more than 50 m. The longer coherent length is shown in the unbalanced Mach-Zender interferometer experiment.

A schematic diagram of the experimental setup using a Mach-Zender interferometer is shown in **Fig. 1**. The path length is 15 m, thus the traveling time of the photon is approximately 50 ns. An AOM is shown in **Fig. 3**; "on" indicates that the incident beam passes through as a zero order beam; "off" indicates that the incident beam diffracts as a first order beam. The photon and wave packets pass through or are diffracted by the AOM: that is, the AOM changes the paths of photons. **Figure 3** shows that photons are not absorbed.

Using a single mode laser, photon 1 and its wave packets are in the paths of a balanced Mach-Zender interferometer, as shown in **Fig. 4 (a)**. That is, photon 1 travels at the speed of light and the wave packets non-locally reach beam splitter 2. When AOM 2 is turned off, as shown in **Fig. 4 (b)**, AOM 2 does not influence photon 1 or its wave packets.

However, when photon 2 is emitted from the laser the wave packet of photon 2 overlaps on the wave packet of photon 1, as shown in **Fig. 4 (c)**. Under a coherent condition, wave packet 2 may interfere with wave packet 1 to modify the interference pattern of photon 1. If the overlap of the wave packets of another photon affects the interference, the operation of AOM 2 simultaneously controls the interference of photon 1 at beam splitter 2. **Figure 5** shows an illustration of the overlapping of the wave packets at beam splitter 2. At beam splitter 2, the wave packet of photon 2 overlaps on the wave packet of photon 1: thus, the interference of photon 1 is affected.

The experimental setup of **Fig. 1** is rather difficult to construct. An experimental setup using optical fibers and couplers is shown in **Fig. 6**. The coupler works as the beam splitter. There is thermal fluctuation on the interference pattern, which will be detected as a sinusoidal fluctuation [5]. However, the experimental setup using optical fiber is easy and feasible.

**Figure 7** shows two expected experimental results of transient phenomena. The path length is 15 m and the traveling time of a photon is 50 ns. The switching time of an AOM is 10 ns, as shown in **Fig. 7 (a)**: the y-axis shows the on and off states of AOM 2. That is, the falling time of the AOM is 10 ns, thus the transient phenomena begin at t=0 ns, and interference disappears at t=50 ns.

The non-local theory predicts a simultaneous interference overlap of wave packets at beam splitter 2. Thus, the outputs of power meters 1 and 2 start changing at t=0, as shown in **Fig. 7 (b)**. The local theory means that the information of a photon is transferred by the photon itself (i.e., photon arrival). The information of AOM 2 is transmitted by the photon itself. Thus, the information travels at the speed of light, *c*, as shown in **Fig. 7 (c)**. The outputs of power meters 1 and 2 start changing at t=50 ns.

The local theory indicates that the wave packet of photon 2 does not affect photon 1, whereas the non-local theory indicates otherwise. At this stage, I will predict the result on the basis of the non-local theory.

Therefore, the wave packet interpretation for the Mach-Zender interferometer results in another hypothesis on non-locality as follows: The overlapping wave packet of another coherent photon affects interference.

## 4. Discussion

Recently, non-locality has received much attention in the field of quantum information, and non-local hidden-variable theory. However, this discussion describes interference, and is carried out under the de Broglie-Bohm picture.

This proposed experiment requires the assumption that a coherent photon interferes with another coherent photon. **Figure 5** shows an illustration of the overlapping of the wave packet of photon 2 over that of photon 1 at beam splitter 2. At beam splitter 2, the wave packet of photon 2 overlaps on the wave packet of photon 1. Thus the wave packet at beam splitter 2 is modified to affect the interference of photon 1. This assumption can be checked experimentally.

The setting of the experimental setup in **Fig. 2** does not appear to cause interference. This is because there are two, separated paths. According to wave packet interpretation, interference seems to occur at beam splitter 2. This conclusion differs from the interpretation, in which the time when a photon reaches beam splitter 2 defines the interference condition. That is, when we see experimental setup in **Fig. 2**, the experimental conditions do not appear to cause interference. However, when wave packets are assumed, interference will be predicted.

## 5. Conclusion

The idea for this discussion occurred when I saw a simulation of Young's double slit experiment. It was a surprise for me. When I carried out Young's double slit experiment, I realized that the simulation shows an Airy pattern, which is shown in the middle range of Young's double slit experiment. When one of the two slits is open, an Airy pattern is seen. At a later time, when the other slit is opened, Young's double slit pattern is observed. A slight transverse motion of a photon transforms the Airy pattern to Young's double slit pattern. Therefore, when Young's double slit experiment is carried out using a single mode laser, I predict that a photon travels at the speed of light, $c$; however, the pattern looks to be non-locally transformed from an Airy pattern to Young's double slit pattern. In this paper, I discussed the experimental setup for measuring the interference speed to obtain the positive results of non-locality. A Mach-Zender interferometer is used to measure the speed of interference. A single mode laser with a large coherent length is used. Although a photon travels at the speed of light, $c$, at this stage, the interference is predicted to be non-local.

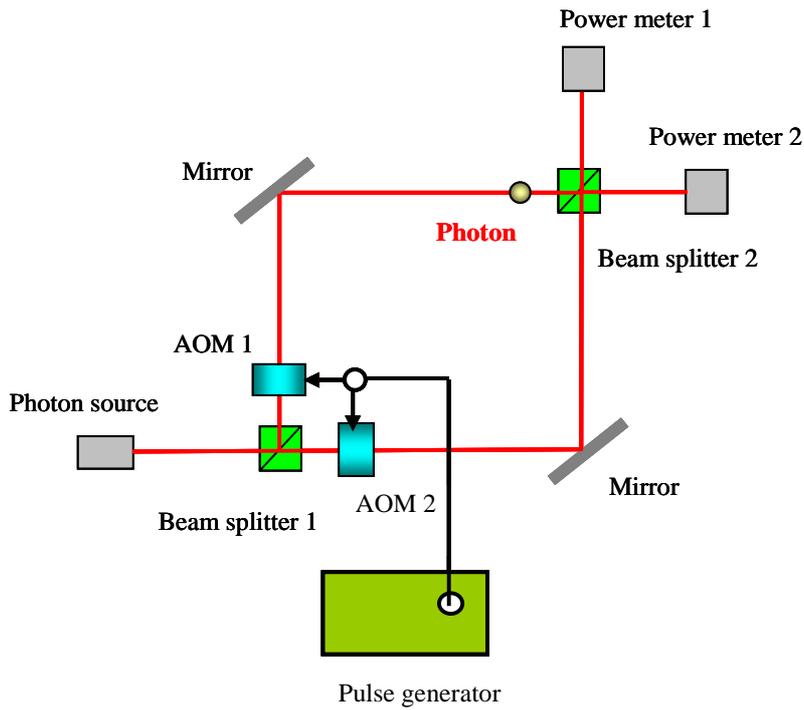

**Fig. 1**　Schematic diagram of the experimental setup for measuring the speed of interference using Mach-Zender interferometer.

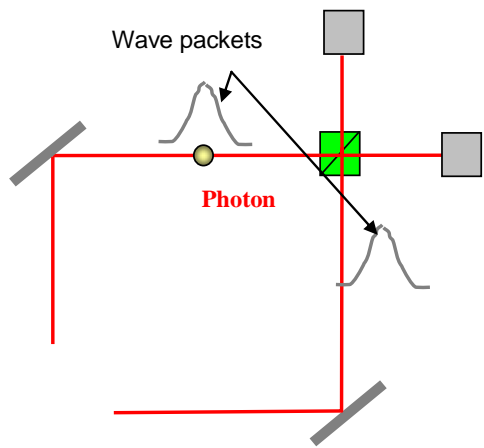

**Fig. 2**　Wave packet interpretation of Mach-Zender interferometer. After the two acousto optic modulators (AOMs) are turned off, the photon and wave packets travel along the paths shown. The photon source is not a coherent laser; thus, the pulse width of the wave packets is small. The interference condition is not controllable by AOM 1 or 2.

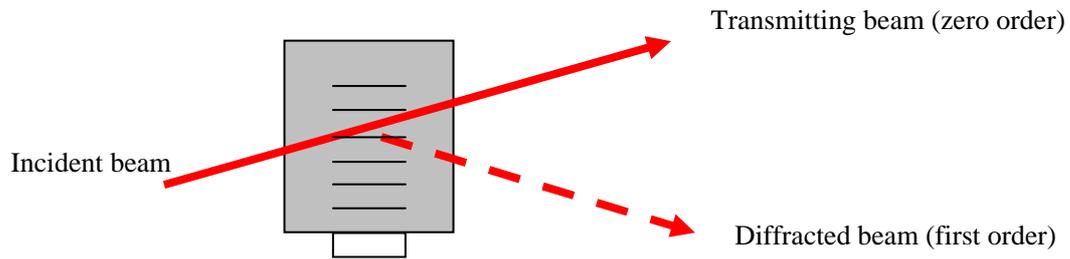

Incident beam

Transmitting beam (zero order)

Diffracted beam (first order)

**Fig. 3** Acousto optic modulator (AOM) changes the paths of photons. Photons are not absorbed.

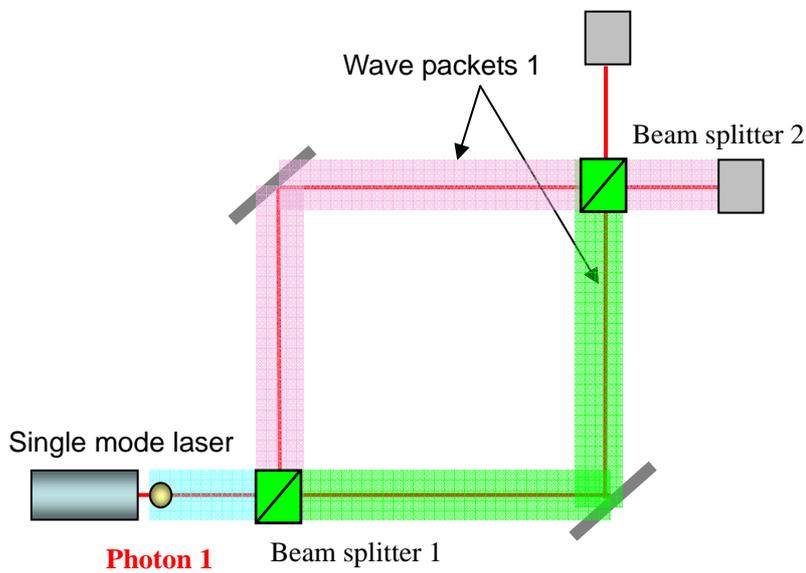

Wave packets 1

Beam splitter 2

Single mode laser

**Photon 1**

Beam splitter 1

**Fig. 4 (a)** Wave packet interpretation of a balanced Mach-Zender interferometer. The interference condition of the photon is defined by the coherent wave packets. The coherent length of the laser is greater than 50 m. Interference appears.

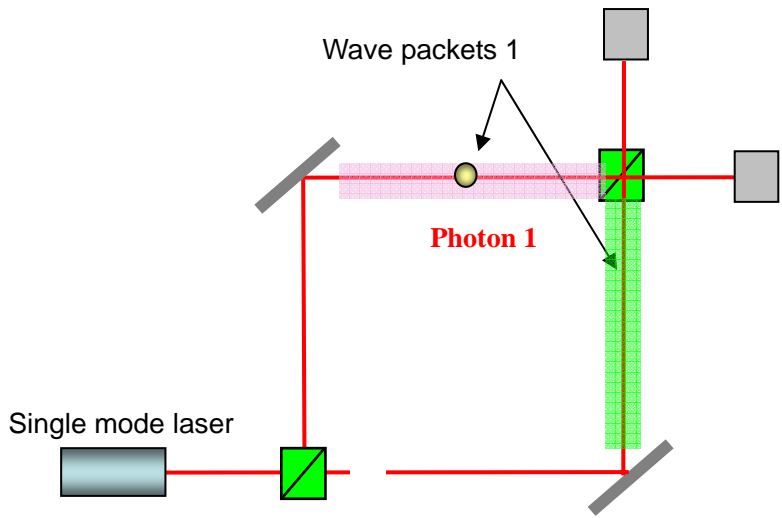

**Fig. 4 (b)** When acousto optic modulator (AOM) 2 is turned off, there are no influences of the control of AOM 2 on photon 1 or its wave packets. Interference appears.

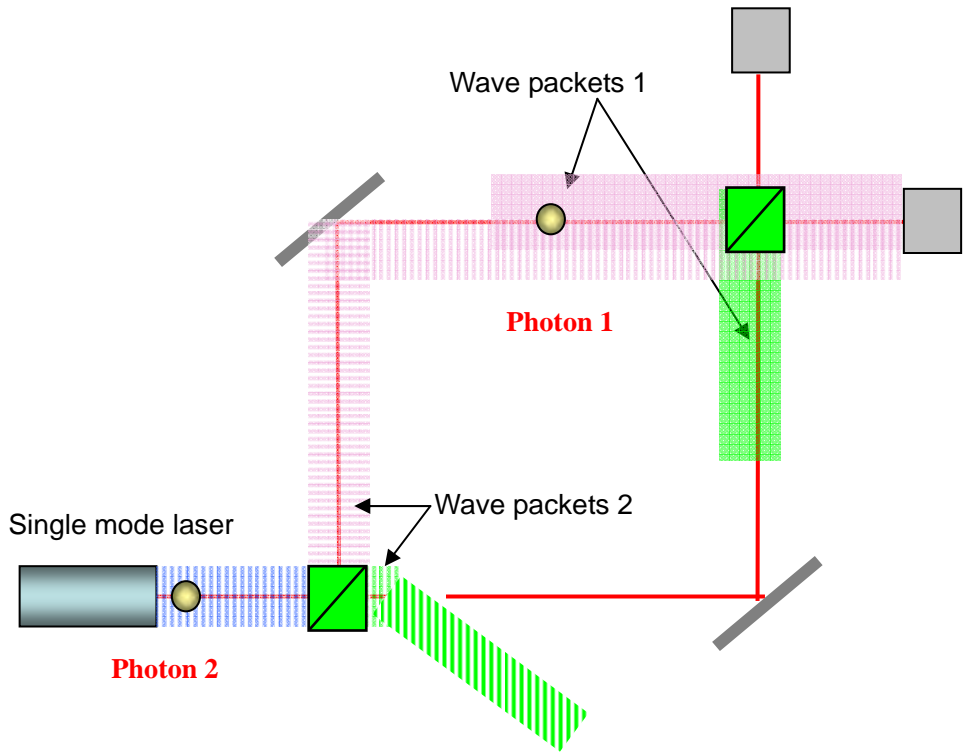

**Fig. 4 (c)** When photon 2 is emitted from the laser, the wave packet of photon 2 overlaps with the wave packet of photon 1. Under a coherent condition, wave packet 2 may interfere with wave packet 1 to modify the interference pattern of photon 1. As a result, acousto optic modulator (AMO) 2 can simultaneously control the interference of photon 1. The interference of photon 1 is affected by the wave packet 2.

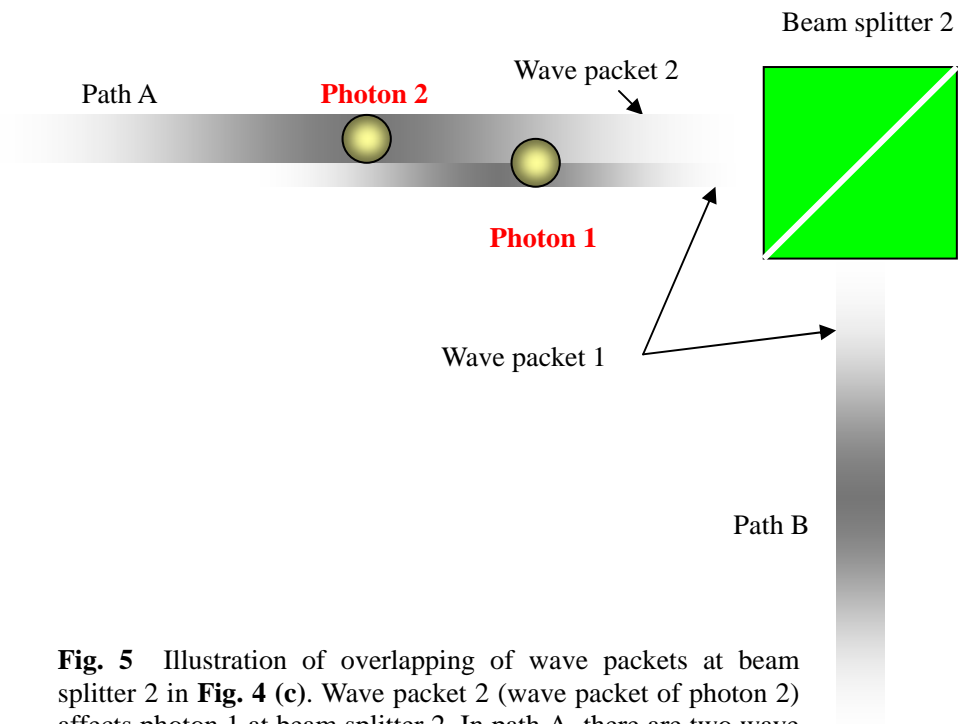

**Fig. 5** Illustration of overlapping of wave packets at beam splitter 2 in **Fig. 4 (c)**. Wave packet 2 (wave packet of photon 2) affects photon 1 at beam splitter 2. In path A, there are two wave packets; in path B there is only one wave packet.

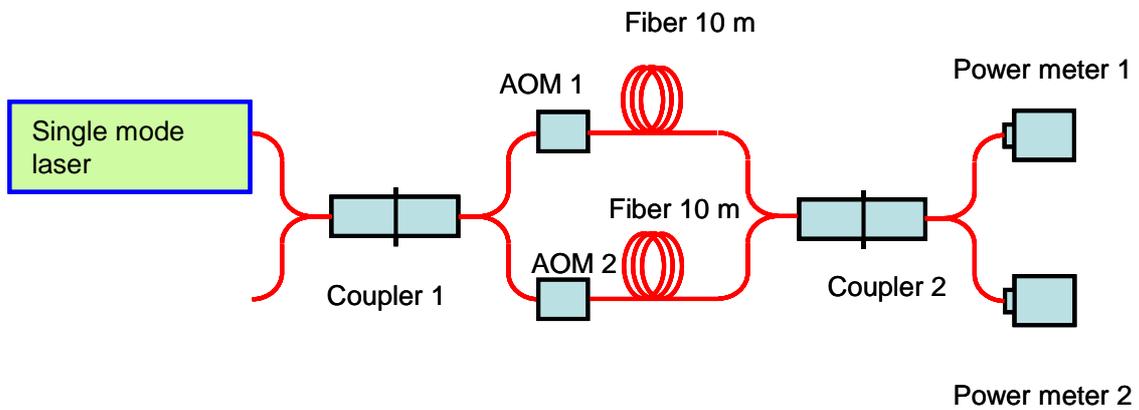

**Fig. 6** Schematic diagram of experimental setup using optical fibers.

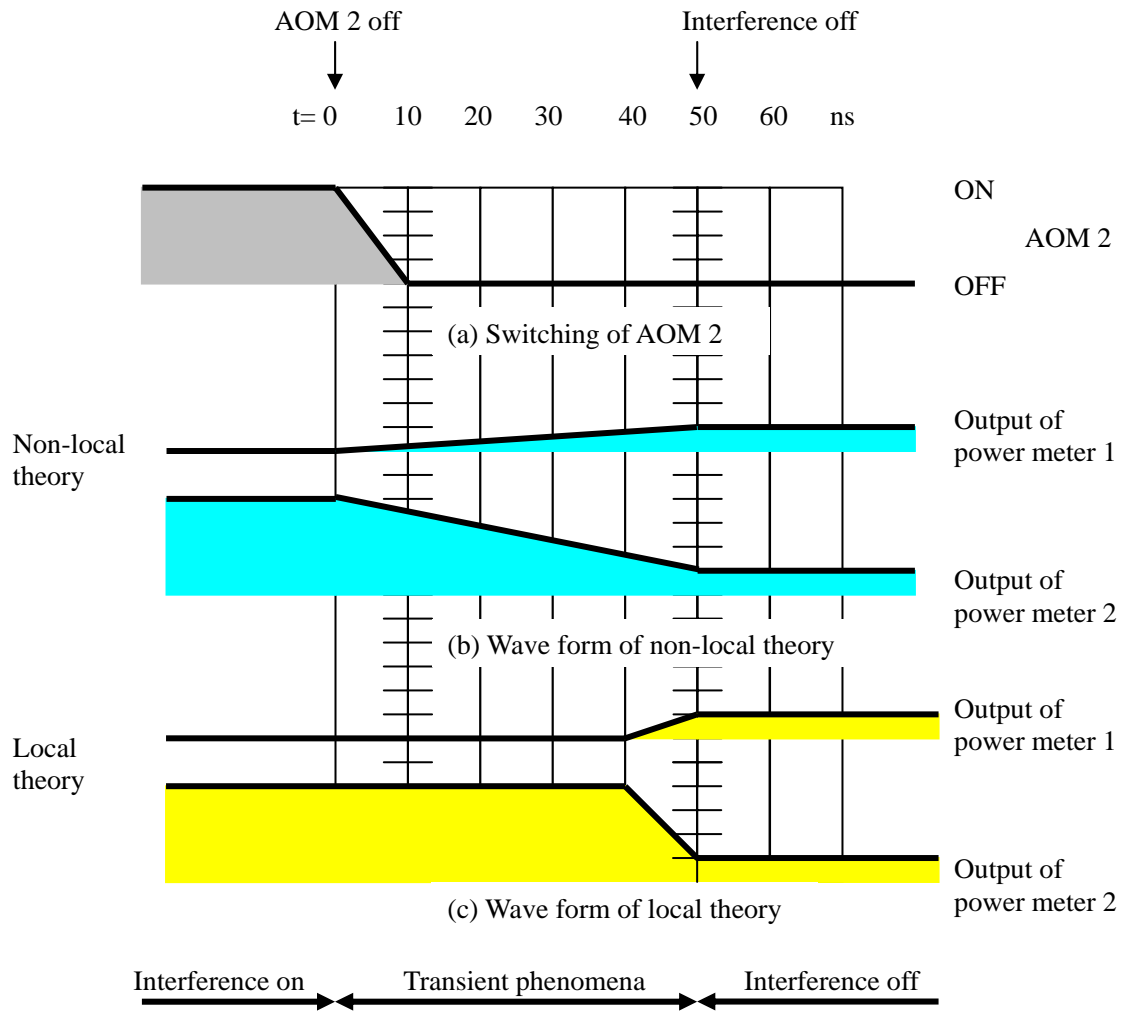

**Fig. 7** Transient phenomena assuming non-local and local theories. Acousto optic modulator (AOM) 2 turns off at t=0, the switching time of the AOM is 10 ns. Transient phenomena are seen between t=0 to t=50 ns. (a) Switching of AOM 2: y-axis shows ON or OFF of AOM 2. (b) Wave form of non-local theory: y-axis shows the output level of power meters 1 and 2. The non-local theory predicts a simultaneous interference change between AOM 2 and power meters 1 and 2. (c) Wave form of local theory: the local theory predicts 50 ns delay of the interference at beam splitter 2. At this stage, I predict non-local theory.